\documentclass[%
 superscriptaddress,
 amsmath,amssymb,
 aps,
 prl,
 floats,
 prl, 
 twocolumn
]{revtex4-1}

\usepackage{graphicx}
\usepackage{dcolumn}
\usepackage{bm}
\usepackage{color}

\usepackage{todonotes}

\begin{document}

\title{
Quantum Critical Point revisited by the Dynamical Mean Field Theory \\
}

\author{Wenhu Xu}
\affiliation{%
 Division of Condensed Matter Physics and Material Science, Brookhaven National Laboratory, Upton, NY 11973
}%

\author{Gabriel Kotliar}%
\affiliation{%
 Division of Condensed Matter Physics and Material Science, Brookhaven National Laboratory, Upton, NY 11973
}%
\affiliation{
 Department of Physics and Astronomy, Rutgers University, Piscataway, NJ 08854
}%

\author{Alexei M. Tsvelik}
\affiliation{%
 Division of Condensed Matter Physics and Material Science, Brookhaven National Laboratory, Upton, NY 11973
}%

\begin{abstract}
Dynamical mean field theory  is used to study  the quantum critical point (QCP) in the doped Hubbard model on a square lattice. The QCP is characterized by a universal scaling form of the self energy and a spin density wave instability at an incommensurate wave vector. The scaling form unifies the low energy kink and the high energy waterfall feature in the spectral function, while the spin dynamics includes both the critical incommensurate and high energy antiferromagnetic paramagnons. We use  the frequency dependent four-point correlation function of spin operators to calculate the momentum dependent correction to the electron self energy. Our results reveal a substantial difference with the calculations based on the Spin-Fermion model which indicates that the  frequency dependence of the the quasiparitcle-paramagnon vertices is an important factor. 

\end{abstract}


\def\bfq{\mathbf{q}}
\def\bfk{\mathbf{k}}
\def\bfQ{\mathbf{Q}}

\maketitle


{\bf Introduction}. The interplay between quasiparticles and bosonic collective modes, in particular in the proximity of a quantum critical point (QCP)~\cite{PhysRevB.14.1165, PhysRevB.48.7183}, is believed to be a driving force behind the rich phase diagram of many correlated electron systems~\cite{keimer2015quantum, coleman2007, RevModPhys.87.855}. The aim of this paper is to explore the connection between the quasiparticles and collective modes in a doped antiferromagnet within the framework of dynamical mean field theory (DMFT)~\cite{RevModPhys.68.13}. This approach  enables us to take a full account of the frequency dependence of the vertex functions and to compare it with the phenomenological approach based on the Spin-Fermion (SF) model \cite{PhysRevLett.84.5608, spin_fermion_abanov} where the vertex functions are replaced by constants. 

With the two-dimensional Hubbard model as a working example, we focus on the instabilities in a correlated metal 
and  explore their wave vector and frequency dependence. In addition to the dynamical susceptibilities we calculate the entire four-point correlation functions and use them to calculate the rainbow diagram correction to the single particle self energy. 
Unlike the SF model, it turns out that the areas of the Brillouin zone most affected by the fluctuations are not located near the Fermi surface (FS) hot spots --- the points on the FS connected by the SDW ordering wave vector ${\bf Q}$. 
The calculation with the full frequency dependent vertex yields a different result: the most significant correction emerges in the antinodal  $(\pi,0)$ and $(0, \pi)$ areas and its frequency dependence is different from the one given by the SF model. 
 
We study the two-dimensional Hubbard Hamiltonian on a square lattice with nearest $t$ and next nearest neighbor hopping $t'$,   
\begin{equation}
H = - \sum_{ \langle i,j \rangle, \sigma } t_{ ij } c_{ i, \sigma }^{ \dagger } c^{ }_{ j, \sigma } + U \sum_{ i } c^{ \dagger }_{ i \uparrow } c^{ }_{ i \uparrow }
c^{ \dagger }_{ i \downarrow }c^{}_{ i \downarrow }. \label{eq:Hubbard} 
\end{equation}
Taking $t = 1$ as the unit for energy and temperature, we set $t' = -0.3$ and the Coulomb interaction $U = 14$. We focus on the band fillings between $n = 0.6$ and $n = 0.9$. To solve the effective impurity problem in the DMFT self-consistent approach  \cite{RevModPhys.68.13} we adopt the continuous time quantum Monte Carlo (CTQMC) method \cite{PhysRevLett.97.076405} as implemented in Ref. \cite{PhysRevB.75.155113}. 

\begin{figure}
 \includegraphics[width=0.45\textwidth]{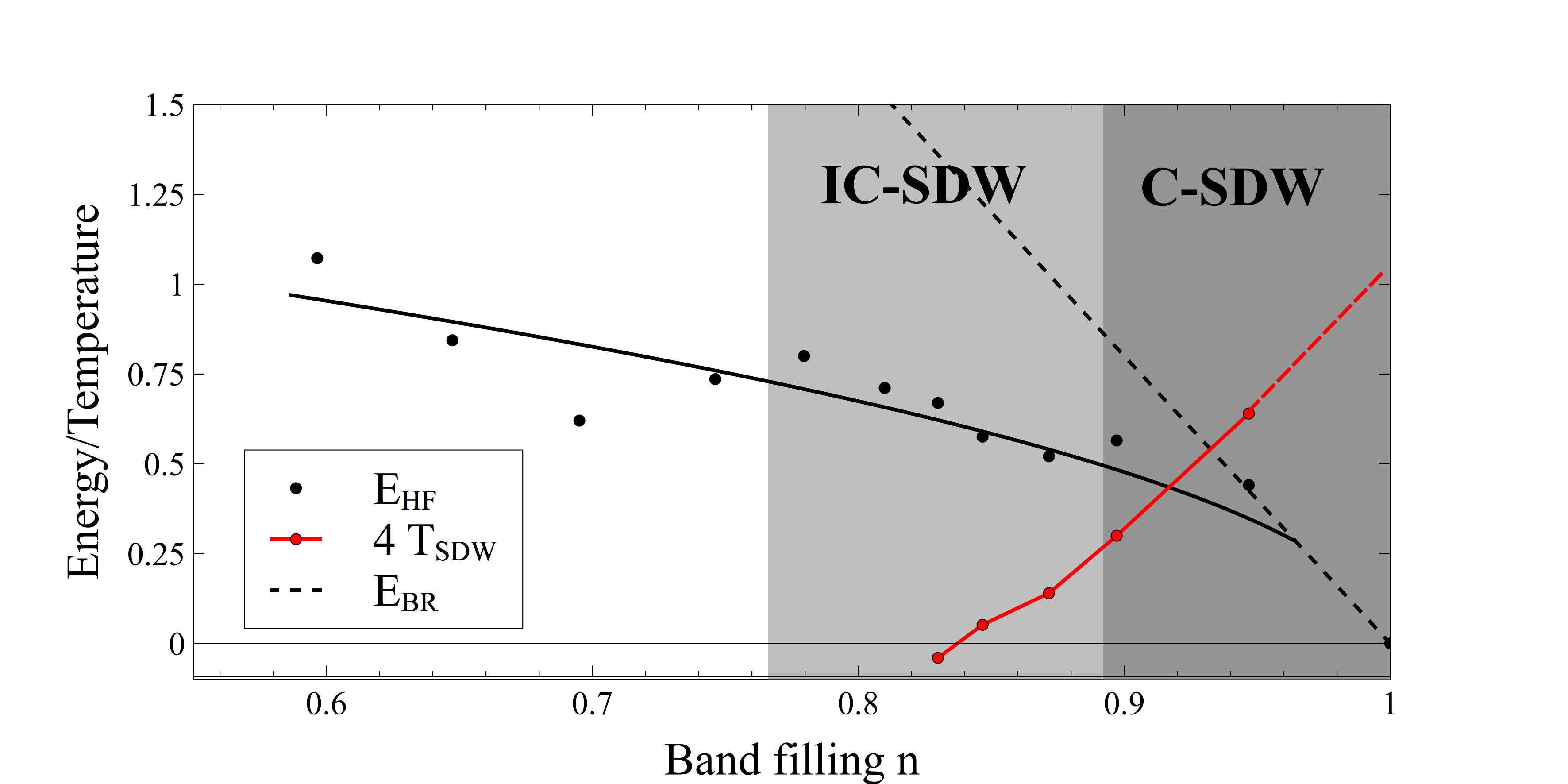}
 \caption{ \label{fig:phase}  The energy scales and the  phase diagram of the doped Hubbard model (\ref{eq:Hubbard}) at $T = 0.02$. $E_{HF}$ is the characteristic energy scale for the frequency dependence of the self energy. The solid line is a guide for the eye. $E_{BR} = (1-n) W$ is  the Brinkman-Rice scale. $T_{SDW}$ is  the SDW transition temperature. Shading highlights regions with commensurate (C) or incommensurate (IC) SDW fluctuations. } 
\end{figure}

Figure~\ref{fig:phase} presents  a summary of the main results. The effective Fermi energy $E_{HF}$ is defined by the energy dependence of quasiparticle (QP) damping rate (see Eq.(2) below). In the region of interest  $E_{HF}$ does not coincide with the Brinkman-Rice scale $T_{BR} = (1-n)W$ ($W$ is the non-interacting bandwidth). 
The interval of $0.76 \lesssim n \lesssim 0.9$ defines the quantum critical region; the SDW transition temperature $T_{SDW}$ vanishes  at $n \simeq 0.84$ with an incommensurate ${\bf Q}$. In this region the frequency dependent part of the Green's function  $\omega -\Sigma(\omega)$ can be fit by a universal function of $\omega/E_{HF}$. Unlike in the conventional Landau--Fermi liquid theory, the  QP residue $Z_{QP}$ in this region is strongly frequency  dependent. Below $n \simeq 0.75$ the critical collective mode disappears. 

 
{\bf The single electron Green's function.} The Hubbard model has been investigated extensively by the DMFT community \cite{RevModPhys.68.13, PhysRevLett.101.186403, PhysRevB.75.045118, PhysRevB.77.033101}, with the most attention focussed on the single electron Green's function $G_{\bfk}(\omega)$ and the density of states. Since the Green's function is also an important ingredient of the SF model, we are compelled to discuss it first.  We start with the local self energy. 

\begin{figure}
 \includegraphics[width=0.45\textwidth]{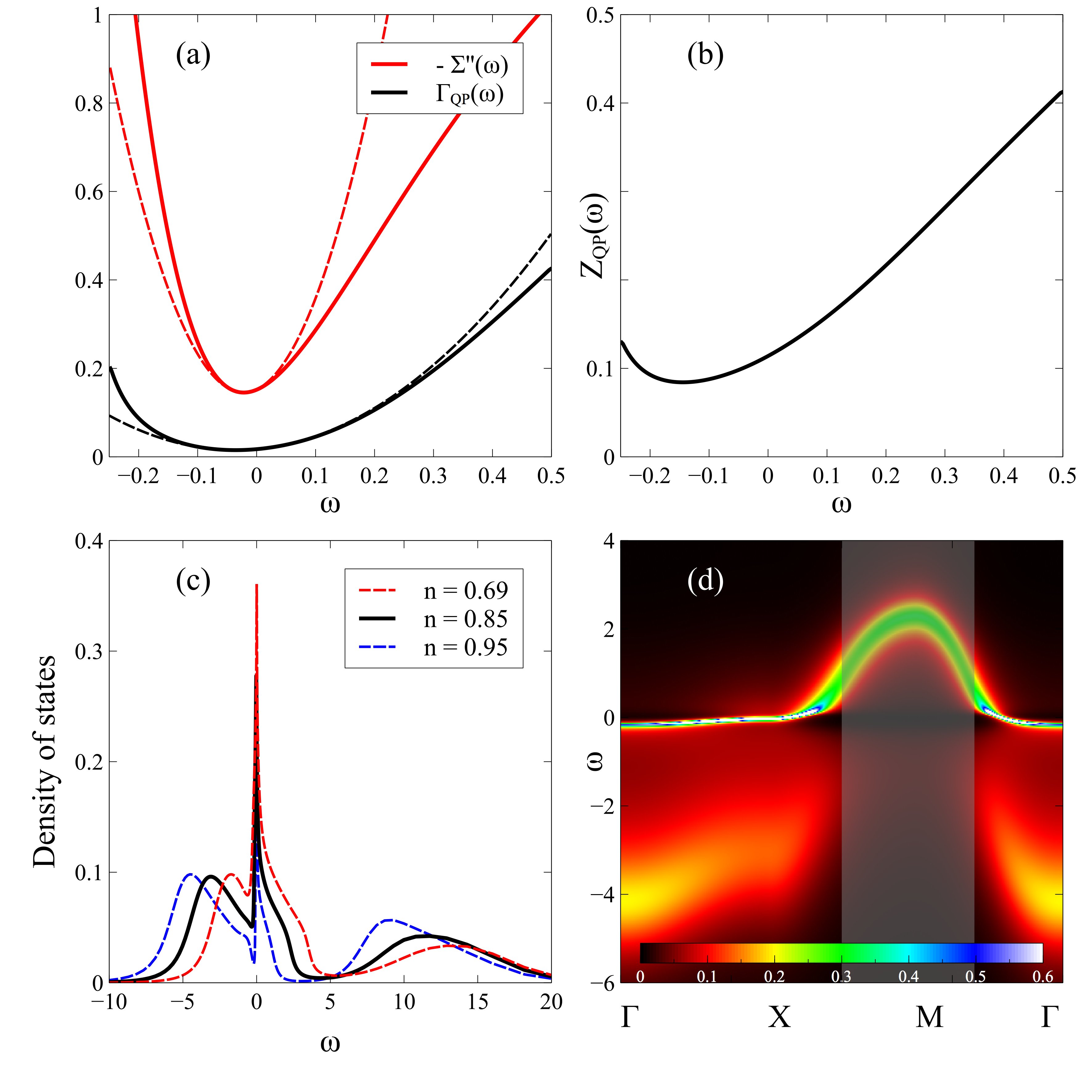}
 \caption{ \label{fig:spectral_n085}  Spectral properties for the band filling $n = 0.85$ and $T = 0.02$. 
 (a) The imaginary part of the self energy $-\Sigma''(\omega)$ and the QP damping $\Gamma_{QP}(\omega)$. The dashed lines represent a quadratic fit in the region $|\omega| \leq 0.05$. 
 (b) The quasiparticle  residue $Z_{QP}(\omega)$. 
 (c) The density of states (DOS). Also shown are the DOS for $n = 0.69$ and $n = 0.95$. 
 (d) The spectral function $A_{\bfk}(\omega) = - G''_{\bfk}(\omega)/\pi$. } 
\end{figure}

\begin{figure}
 \includegraphics[width=0.45\textwidth]{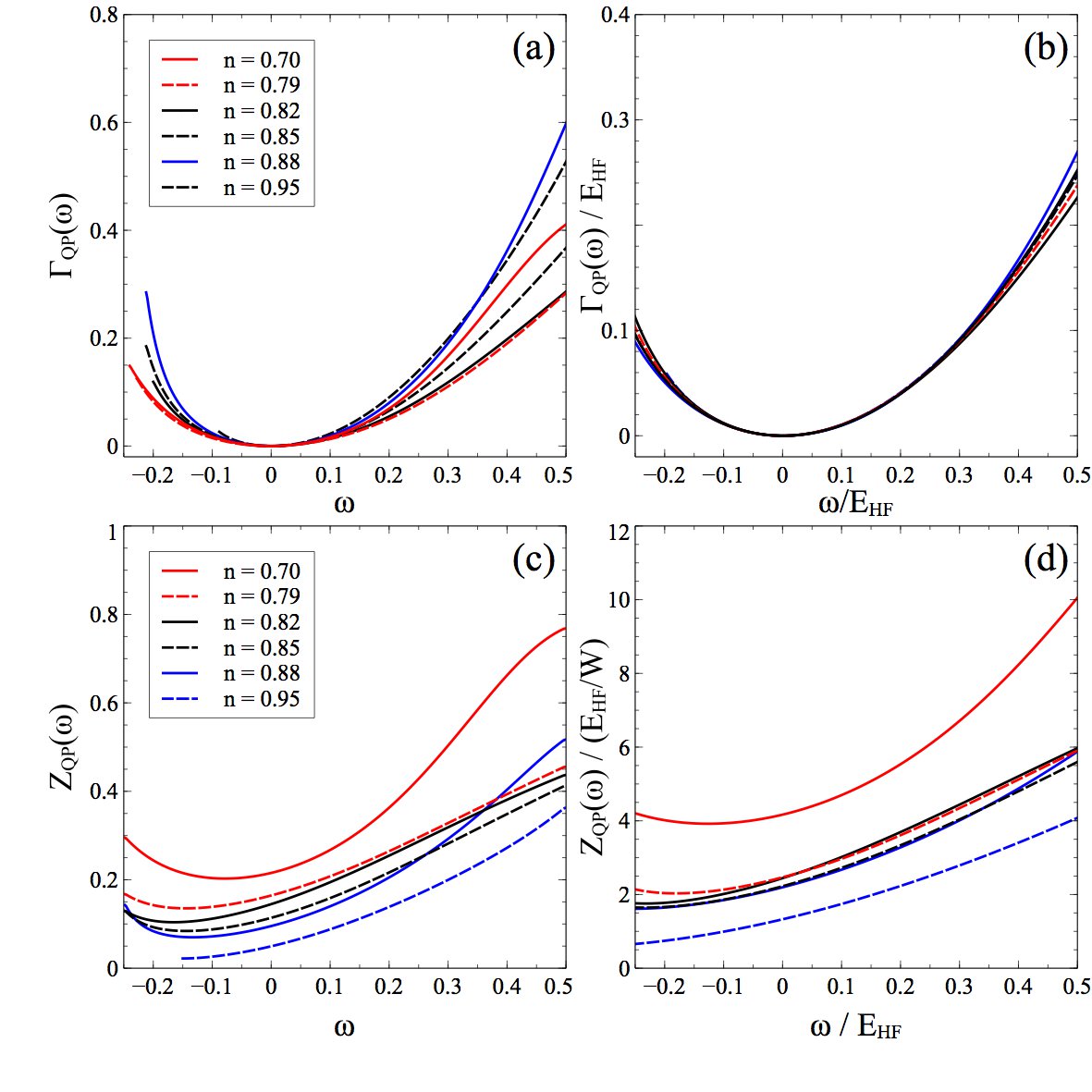}
 \caption{ \label{fig:scaling}  (a) The QP damping rate; (b) the QP damping rescaled by the renormalized Fermi energy $E_{HF}$ of the hidden Ferm liquid; (c) the QP residue; (d) the QP residue rescaled by $E_{HF}$.  } 
\end{figure}

 The physical meaning of $E_{HF}$ can be grasped from Figure \ref{fig:spectral_n085}(a): it is the  energy scale below which a band of heavy fermions is formed. These fermions can be considered as QPs,  as at small frequencies $|\omega| < E_{HF}$ their damping rate is quadratic in frequency: $\Gamma_{QP}(\omega) = -Z_{QP}\Sigma'' \sim \omega^2$. A strong frequency dependence of $Z_{QP} = (1 - d\Sigma'(\omega)/d\omega)^{-1}$, as shown in Figure~\ref{fig:spectral_n085}(b), helps the QPs to remain well defined in the entire frequency interval below $E_{HF}$. Notice that the self energy $\Sigma(\omega)$ itself is quadratic only in a very narrow range of frequencies: $|\omega| \lesssim 0.05$. The dashed lines in Figure~\ref{fig:spectral_n085}(a) are quadratic  fits of $\Gamma(\omega)$ and $\Sigma''(\omega)$, suggesting the robustness of the QPs far beyond $|\omega| = 0.05$. This  strong energy/temperature dependence in $Z_{QP}$  has given rise to the concept of hidden Fermi liquid (HFL)\cite{PhysRevLett.111.036401, PhysRevLett.110.086401, PhysRevLett.113.246404}, in which the linear resistivity and other anomalous transport properties of correlated metals are consequences of the strong temperature dependence of $Z_{QP}$, but the QPs remain well defined such that transport can be treated  within the Boltzmann framework. 

In our effort to determine the degree of universality present in the model (\ref{eq:Hubbard})  we fit the frequency dependent part of the Green's function self energy by the scaling form:
\begin{equation}
\omega - \tilde{ \Sigma } (\omega) = W \times g(\omega/E_{HF}), \label{eq:scaling}
\end{equation}
where $\tilde{ \Sigma } (\omega) = \Sigma( \omega ) - \Sigma (\omega=0)$ and $W$ is the non-interacting bandwidth. As illustrated by Figure \ref{fig:scaling}(b) and (d),  this scaling works very well in the vicinity of the QCP, namely, for the interval $0.79 <n < 0.88$ and $|\omega| < E_{HF}/2$. It is noteworthy that near to the QCP $E_{HF}$ deviates from the Brinkman-Rice scale $T_{BR}$. Thus the  proposed scaling form  differs from the scaling form used in the  recent renormalization group study~\cite{Mai2016}.

 A strong  $\omega$-dependence of  $Z_{QP}$ leads to strong momentum dependency of  the QP residue. The latter fact enables us to explain the famous ``waterfall'' phenomenon observed in angular-resolved photoemission experiments \cite{PhysRevLett.98.067004, PhysRevLett.99.237002, PhysRevLett.109.066404, PhysRevX.2.041012}. This phenomenon amounts to vanishing of the spectral weight in the lower Hubbard band  in a particular region of the Brillouin zone. In the strong coupling limit ($U \rightarrow \infty$), the high energy excitations with double occupancy (the upper Hubbard band) carry spectral weight of $n/2$~\cite{Hubbard238}. Therefore the combined spectral weight for the hidden QPs and the lower Hubbard band is $1-n/2$ (Figure~\ref{fig:spectral_n085}(c)). In the shaded region of Figure~\ref{fig:spectral_n085}(d) the hidden QPs with $\omega > 0.5$ exhaust all available spectral weight leaving nothing for the states below the chemical potential.  Approaching the Fermi surface $Z_{QP}$ decreases,  giving rise  to the kink in the QP dispersion and the emergence of  the incoherent continuum. The incoherent continuum at $\omega < -2$ is placed around the bare band dispersion and is connected to the QP band by the spectral intensity in the vertical direction (the ``waterfall''). 
 
  
 We see that  the high energy waterfall feature, along with the low energy ``kink'' feature in the QP band, is a consequence of the $\omega$-dependent self energy which is characterized by a single energy scale $E_{HF}$. As far as we are aware, this connection has not been addressed in previous works~\cite{PhysRevLett.99.237001, PhysRevB.78.134519, PhysRevLett.100.066406, PhysRevB.78.212504, PhysRevB.82.134505, PhysRevB.83.235113}. 


{\bf  Lattice susceptibilities in the spin channel.}
Computing the lattice susceptibility of  bosonic modes within DMFT requires an extra effort~\cite{RevModPhys.68.13}. Firstly, one needs to determine the irreducible vertex in the spin channel $\Gamma^{S}(i\nu, i\nu')_{i\Omega}$. It  is computed by the Bethe-Salpeter equation:
\begin{multline}
\left[ \chi_{loc}^{S}(i\nu, i\nu')_{ i\Omega } \right]^{-1} \\
=  \left[ \chi_{loc}^{0}(i\nu, i\nu')_{ i\Omega } \right]^{-1}  + 
 \frac{1}{\beta^2} \Gamma^{S}(i\nu, i\nu')_{ i\Omega }, \label{eq:BS_chi_vv_locOm}
\end{multline}
where $\chi^{S}_{loc}(i\nu, i\nu')_{ i\Omega }$ is the local two-particle correlation function  and $\chi^0_{loc}(i\nu, i\nu')_{ i\Omega } = -\beta G_{loc}(i\nu) G_{loc}(i\nu+i\Omega)\delta_{\nu\nu'}$ is the local polarization bubble. $G_{loc}(i\nu)$ is the local Green's function fully dressed by the self energy. $\chi^{S}_{loc}(i\nu, i\nu')_{ i\Omega }$ and $G_{loc}(i\nu)$ are sampled by the CTQMC solver.

The lattice ($\bfq$-dependent) two-particle correlation function $\chi^{S}(i\nu, i\nu')_{\bfq{}, i\Omega}$ is constructed from $\Gamma^{S}(i\nu, i\nu')_{i\Omega}$ and the $\bfq$-dependent polarization bubble $\chi_0(i\nu, i\nu')_{\bfq{}, i\Omega} = -\beta \sum_{\bfk{}} G_{\bfk}(i\nu) G_{\bfk + \bfq}(i\nu+i\Omega)\delta_{\nu\nu'}$, 
\begin{multline}
 \left[ \chi^{S}(i\nu, i\nu')_{\bfq{}, i\Omega} \right]^{-1}  \\
 =  \left[ \chi_0(i\nu, i\nu')_{\bfq{}, i\Omega} \right]^{-1}  + 
 \frac{1}{\beta^2} \Gamma^{S}(i\nu, i\nu')_{i\Omega},  \label{eq:BS_chi_vv_qOm}
\end{multline}
The dynamical susceptibility in the Matsubara frequency domain is then calculated by closing the fermionic frequencies, 
$\chi^{S}(\bfq, i\Omega)  = \frac{1}{\beta^2}\sum_{\nu, \nu'} \chi^{S}(i\nu, i\nu')_{ \bfq, i\Omega }$.
Finally, we fit $\chi(\bfq, i\Omega)$ by the damped model expression to determine the resonance energy $\Omega^S(\bfq)$ and the damping rate $\gamma^S(\bfq)$, 
\begin{equation}
\chi^{S}(\bfq, i\Omega) = \frac{ \chi^{S}(\bfq) \Omega^{S}(\bfq)^{2} }{ \Omega^{S}(\bfq)^2 + |\Omega|^2 + \gamma^{S}(\bfq) |\Omega|  } \label{eq:chiqw}
\end{equation}
where $\chi(\bfq)$ is the static susceptibility at generic $\bfq$. The analytical continuation to real frequencies is straightforward by taking $i\Omega \rightarrow \Omega+ i0^{+}$ in the damped model. 

\begin{figure}
 \includegraphics[width=0.45\textwidth]{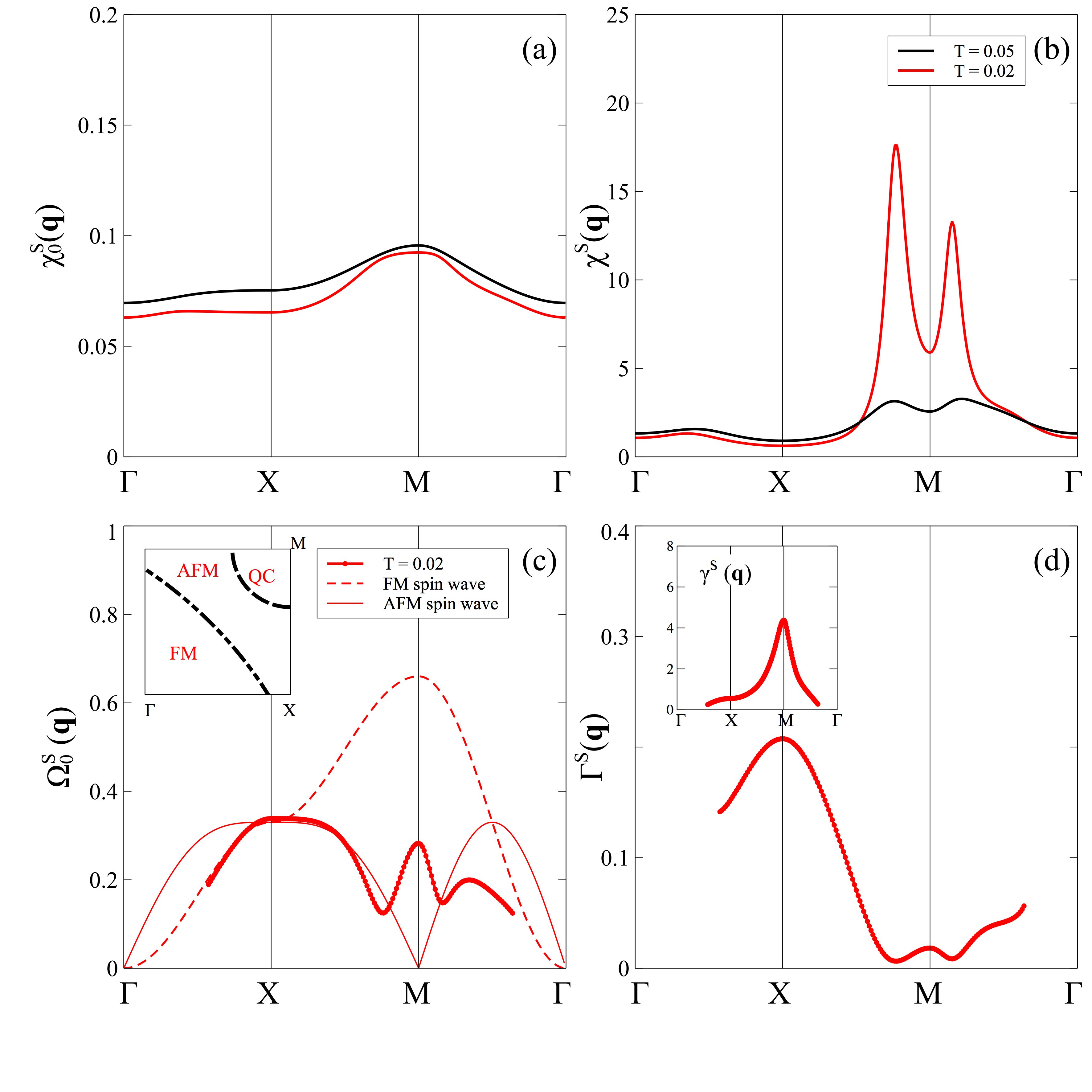}
  \caption{ \label{fig:chiQ}  (a)  The bare static lattice susceptibility.  (b) The full static lattice susceptibility. (c) The resonance energy $\Omega^{S}(\bfq)$. Inset of (c): Partition of $\bfq$-space by the nature of spin fluctuations. (d) The relaxation rate in the overdamped region near the M point, $\Gamma^{S}(\bfq) = (\Omega^{S}(\bfq))^2/\gamma^{S}(\bfq)$. Inset of (d): the damping rate $\gamma^{S}(\bfq)$. The band filling is taken as $n = 0.85$.  }
\end{figure}

Figure~\ref{fig:chiQ}(a) shows the ``bare'' static susceptibility $\chi_{0}({\bfq})$ (the polarization bubble). Without the vertex correction, $\chi_{0} (\bfq)$ is only weakly dependent on $\bfq$ and temperature, suggesting that the instability does not originate from  the particle-hole excitations at the Fermi surface. Figure~\ref{fig:chiQ} (b) presents the full static susceptibilities $\chi^{S}(\bfq)$. $\chi^{S}(\bfq)$ is significantly enhanced for all $\bfq$'s, and exhibits qualitatively different temperature dependencies near the $\Gamma$ and M points in $\bfq$-space. Near the $\Gamma$ point, $\chi^{S}(\bfq)$ is only weakly temperature dependent, while near the $M$ point, $\chi^{S}({\bfq})$ exhibits a dramatic increase at low temperatures, in particular for the wave vectors sitting on the ring centered at the M point. The $\bfq$'s sitting on this ring are not equally critical. Extrapolating  $1/\chi^{S} (\bfq)$ as a function of $T$ to $0$ we can pick $\bfQ$ with the highest transition temperature $T_{SDW}$ as the SDW ordering wave vector.  Figure~\ref{fig:phase} shows the variation of $T_{SDW}$ and $\bfQ$ with band filling. $\bfQ = (\pi, \pi)$ for $n \geq 0.9$ and is incommensurate for $ n \leq 0.9$. For instance,  for band filling $n = 0.85$, we have $\bfQ \simeq (\pi \pm 0.2\pi, \pi)$ and $(\pi, \pi \pm 0.2\pi)$. The QCP for the SDW order is located at $n \simeq 0.84$. 

The damped model (Eq.~(\ref{eq:chiqw})) reveals the partition of the spin excitations in $\bfq$-space. As shown in Figure~\ref{fig:chiQ}(c), near the $\Gamma$ point the resonance energy $\Omega^{S}_0$ follows the dispersion of the ferromagnetic (FM) spin wave. Passing the $X$ point and approaching  the $M$ point, $\Omega^S_{0}(\bfq)$ traces the antiferromagnetic (AFM) spin wave before entering the critical ring centered on the $M$ point, where $\Omega^S_{0}(\bfq)$ develops a ``mexican hat'' shape. The damping rate $\gamma^{S}(\bfq)$ (inset of Figure~\ref{fig:chiQ}(d)) is peaked at the M point, suggesting that the FM paramagnons near $\Gamma$ are only slightly damped while the critical SDW paramagnons are overdamped and characterized by the relaxation rate $\Gamma^{S}(\bfq) \equiv (\Omega^{S}(\bfq))^{2}/\gamma^{S}(\bfq)$. The coexistence of heavily damped AFM fluctuations at high energy and incommensurate critical paramagnons at low energy, along with the sign of FM fluctuations, resembles the spin dynamics in the normal state of cuprate superconductors as measured by neutron and resonant inelastic X-ray scattering (RIXS) experiments~\cite{hinkov2007spin, vignolle2007two, PhysRevB.78.132503, xu2009testing, le2011intense, jia2014persistent}.  

\begin{figure}
 \includegraphics[width=0.45\textwidth]{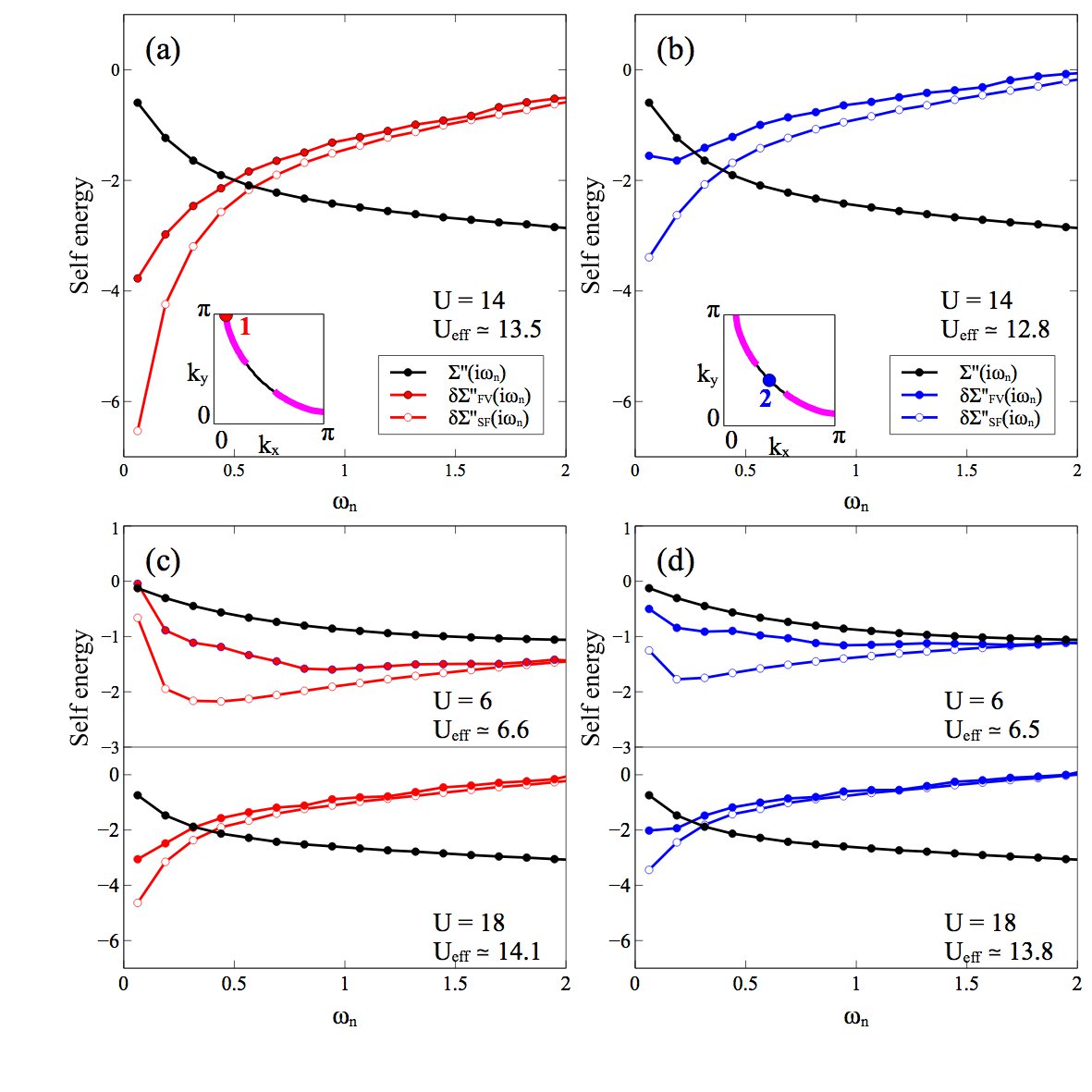}
 \caption{ \label{fig:sigma}  Imaginary part of the self energy calculated with the full vertex (FV) functions and with spin-fermion (SF) models for antinodal point (\textcolor{red}{$\mathbf{1}$}) and nodal point (\textcolor{blue}{$\mathbf{2}$}) for $U=14$ in (a) and (b), $U=6, 18$ in (c) and (d). Non-local corrections to the DMFT self energy, $\delta\Sigma_{FV/SF}(\bfk, i\omega) = \Sigma_{\text{FV}/\text{SF}}(\bfk, i\omega) - \Sigma(i\omega)$ are presented with the local self energy $\Sigma(i\omega)$. Also shown are the values of the effective Coulomb interaction $U_{eff}$ for the SF model (Eq. (\ref{eq:Sig_sf})). } 
\end{figure}

With the four-point correlation function and vertex function at hand we can calculate the non-local self energy due to the emission of paramagnons. Although a fully self-consistent vertex calculation requires an extended DMFT framework, such as the dynamical vertex approximation~\cite{PhysRevB.75.045118} or the dual fermion approach~\cite{PhysRevB.77.033101, PhysRevB.79.045133}, the leading order in the $\bfq$-dependent full vertex (FV) functions, $\Gamma^{S/C}(i\omega, i\omega')_{\bfq, i\Omega}$, provides us with a convenient way to examine the effect of incommensurate paramagnons, 
\begin{eqnarray}
&\Sigma_{\text{FV}}&(\bfk, i\omega) \nonumber \\
{} &=& \frac{1}{2}Un + \frac{U}{2\beta^3}\sum_{\omega',\Omega,\bfq} \chi_0(i\omega, i\omega')_{\bfq, i\Omega}G_{\bfk+\bfq}(i\omega+i\Omega) \nonumber \\
{}&&\qquad\times\left[ 3 \Gamma^{S}(i\omega, i\omega')_{\bfq, i\Omega} - \Gamma^{C}(i\omega, i\omega')_{\bfq, i\Omega} \right. \nonumber \\
{}&& \left. \qquad \qquad + \Gamma^{C}_{loc}(i\omega, i\omega')_{i\Omega} - \Gamma^{S}_{loc}(i\omega, i\omega')_{i\Omega} \right].
\label{eq:Sig_fv}
\end{eqnarray}
$\Gamma^{S/C}(i\omega, i\omega')_{\bfq, i\Omega}$ are calculated from the irreducible local vertex functions $\Gamma^{S/C}(i\omega, i\omega')_{i\Omega}$ via 
\begin{multline}
\left[ \frac{1}{\beta^2}\Gamma^{S/C}(i\omega, i\omega')_{\bfq, i\Omega} \right]^{-1}\\ 
= \left[ \frac{1}{\beta^2}\Gamma^{S/C}(i\omega, i\omega')_{i\Omega}\right]^{-1} - \chi_0(i\omega, i\omega')_{\bfq, i\Omega} .
\end{multline}

Instead of isolated hot spots, the soft paramagnon fluctuations connect continuous segments of the Fermi surface, forming a hot region marked by purple in the inset of Figure~\ref{fig:sigma}(a) and (b). We depict the local self energy $\Sigma^{''}(i\omega)$ and the non-local correction $\delta\Sigma^{''}_{FV}(\bfk, i\omega) = \Sigma^{''}_{FV}(\bfk, i\omega) - \Sigma^{''}(i\omega)$ calculated by Eq. (\ref{eq:Sig_fv}) for $\bfk$'s at the antinodal point(\textcolor{red}{$\mathbf{1}$}) and the nodal point (\textcolor{blue}{$\mathbf{2}$}) in Figure~\ref{fig:sigma}. For comparison, we also show the non-local correction calculated with the dynamical susceptibility, $\delta \Sigma^{''}_{SF}(\bfk, i\omega) = \Sigma^{''}_{SF}(\bfk, i\omega) - \Sigma^{''}(i\omega)$, where
\begin{multline}
\Sigma_{\text{SF}}({\bf k}, i\omega) = \\
 \frac{U_{eff}^2}{\beta}\sum_{\Omega, \bfq}\chi^S(\bfq, i\Omega)G_{{\bf q} + {\bf k}}(i\Omega + i\omega), \label{eq:Sig_sf}
\end{multline}
as is done in the SF model~\cite{PhysRevLett.84.5608, spin_fermion_abanov}. The effective Coulomb interaction $U_{eff}$ is chosen to get the best fit to $\Sigma^{''}(i\omega)$ at high frequency. 

Figures~\ref{fig:sigma} shows the results for $U=6, 14$, and $18$. At high energies, the FV and the SF model calculations lead to essentially the same results for all these values of $U$. The diminishing high frequency tails of $\delta\Sigma_{FV/SF}(\bfk, i\omega)$ at both antinodal and nodal point indicate that the non-local correction is significant only at low energy. This provides a justification for the entire approach which uses the local DMFT as its starting point.

At large $U=14, 18$ the difference between the FV and SF results becomes quite pronounced at small energies. Although in both FV and SF model the self energy shows strong frequency dependence, the momentum dependence is different. As expected, we get a significant energy dependence at the antinode, but the SF model which does not take into account the frequency dependence of the vertex function, overestimates it. At the nodal point, the energy dependence of $\delta\Sigma^{''}_{FV}(\bfk, i\omega)$ resembles that of a Fermi liquid, while the SF model gives strong scattering.  It should also be noted that to achieve a convergence between the FV and the SF at high energy we needed to adopt a very large value for $U_{eff}$ which makes the entire approach based on the Spin Fermion model questionable.


{\bf Conclusions.} 
Using the example of the Hubbard model, we have demonstrated how the standard DMFT procedure can be augmented by the inclusion of corrections from interactions of quasiparticles with collective excitations. Our calculation points to the pivotal role of incommensurate critical paramagnons in a doped antiferromagnet, making contact with recent neutron scattering and RIXS measurements~\cite{hinkov2007spin, vignolle2007two, PhysRevB.78.132503, xu2009testing, le2011intense}. Such corrections become significant near the QCP, as predicted by the phenomenological SF model. However, our calculations which take into account frequency dependence of the interaction vertices indicate that the SF model overestimates these corrections and therefore may require a revision.
 
{\bf Acknowledgements}
\label{sec:acknowledge}
The authors are supported by Center for Computational Design of Functional Strongly Correlated Materials and Theoretical Spectroscopy under DOE grant DE-FOA-0001276.

\bibliography{reference}

\begin{thebibliography}{35}%
\makeatletter
\providecommand \@ifxundefined [1]{%
 \@ifx{#1\undefined}
}%
\providecommand \@ifnum [1]{%
 \ifnum #1\expandafter \@firstoftwo
 \else \expandafter \@secondoftwo
 \fi
}%
\providecommand \@ifx [1]{%
 \ifx #1\expandafter \@firstoftwo
 \else \expandafter \@secondoftwo
 \fi
}%
\providecommand \natexlab [1]{#1}%
\providecommand \enquote  [1]{``#1''}%
\providecommand \bibnamefont  [1]{#1}%
\providecommand \bibfnamefont [1]{#1}%
\providecommand \citenamefont [1]{#1}%
\providecommand \href@noop [0]{\@secondoftwo}%
\providecommand \href [0]{\begingroup \@sanitize@url \@href}%
\providecommand \@href[1]{\@@startlink{#1}\@@href}%
\providecommand \@@href[1]{\endgroup#1\@@endlink}%
\providecommand \@sanitize@url [0]{\catcode `\\12\catcode `\$12\catcode
  `\&12\catcode `\#12\catcode `\^12\catcode `\_12\catcode `\%12\relax}%
\providecommand \@@startlink[1]{}%
\providecommand \@@endlink[0]{}%
\providecommand \url  [0]{\begingroup\@sanitize@url \@url }%
\providecommand \@url [1]{\endgroup\@href {#1}{\urlprefix }}%
\providecommand \urlprefix  [0]{URL }%
\providecommand \Eprint [0]{\href }%
\providecommand \doibase [0]{http://dx.doi.org/}%
\providecommand \selectlanguage [0]{\@gobble}%
\providecommand \bibinfo  [0]{\@secondoftwo}%
\providecommand \bibfield  [0]{\@secondoftwo}%
\providecommand \translation [1]{[#1]}%
\providecommand \BibitemOpen [0]{}%
\providecommand \bibitemStop [0]{}%
\providecommand \bibitemNoStop [0]{.\EOS\space}%
\providecommand \EOS [0]{\spacefactor3000\relax}%
\providecommand \BibitemShut  [1]{\csname bibitem#1\endcsname}%
\let\auto@bib@innerbib\@empty
\bibitem [{\citenamefont {Hertz}(1976)}]{PhysRevB.14.1165}%
  \BibitemOpen
  \bibfield  {author} {\bibinfo {author} {\bibfnamefont {J.~A.}\ \bibnamefont
  {Hertz}},\ }\href {\doibase 10.1103/PhysRevB.14.1165} {\bibfield  {journal}
  {\bibinfo  {journal} {Phys. Rev. B}\ }\textbf {\bibinfo {volume} {14}},\
  \bibinfo {pages} {1165} (\bibinfo {year} {1976})}\BibitemShut {NoStop}%
\bibitem [{\citenamefont {Millis}(1993)}]{PhysRevB.48.7183}%
  \BibitemOpen
  \bibfield  {author} {\bibinfo {author} {\bibfnamefont {A.~J.}\ \bibnamefont
  {Millis}},\ }\href {\doibase 10.1103/PhysRevB.48.7183} {\bibfield  {journal}
  {\bibinfo  {journal} {Phys. Rev. B}\ }\textbf {\bibinfo {volume} {48}},\
  \bibinfo {pages} {7183} (\bibinfo {year} {1993})}\BibitemShut {NoStop}%
\bibitem [{\citenamefont {Keimer}\ \emph {et~al.}(2015)\citenamefont {Keimer},
  \citenamefont {Kivelson}, \citenamefont {Norman}, \citenamefont {Uchida},\
  and\ \citenamefont {Zaanen}}]{keimer2015quantum}%
  \BibitemOpen
  \bibfield  {author} {\bibinfo {author} {\bibfnamefont {B.}~\bibnamefont
  {Keimer}}, \bibinfo {author} {\bibfnamefont {S.}~\bibnamefont {Kivelson}},
  \bibinfo {author} {\bibfnamefont {M.}~\bibnamefont {Norman}}, \bibinfo
  {author} {\bibfnamefont {S.}~\bibnamefont {Uchida}}, \ and\ \bibinfo {author}
  {\bibfnamefont {J.}~\bibnamefont {Zaanen}},\ }\href@noop {} {\bibfield
  {journal} {\bibinfo  {journal} {Nature}\ }\textbf {\bibinfo {volume} {518}},\
  \bibinfo {pages} {179} (\bibinfo {year} {2015})}\BibitemShut {NoStop}%
\bibitem [{\citenamefont {Coleman}(2007)}]{coleman2007}%
  \BibitemOpen
  \bibfield  {author} {\bibinfo {author} {\bibfnamefont {P.}~\bibnamefont
  {Coleman}},\ }\enquote {\bibinfo {title} {Heavy fermions: Electrons at the
  edge of magnetism},}\ in\ \href@noop {} {\emph {\bibinfo {booktitle}
  {Handbook of Magnetism and Advanced Magnetic Materials}}}\ (\bibinfo
  {publisher} {John Wiley and Sons, Ltd},\ \bibinfo {year} {2007})\BibitemShut
  {NoStop}%
\bibitem [{\citenamefont {Dai}(2015)}]{RevModPhys.87.855}%
  \BibitemOpen
  \bibfield  {author} {\bibinfo {author} {\bibfnamefont {P.}~\bibnamefont
  {Dai}},\ }\href {\doibase 10.1103/RevModPhys.87.855} {\bibfield  {journal}
  {\bibinfo  {journal} {Rev. Mod. Phys.}\ }\textbf {\bibinfo {volume} {87}},\
  \bibinfo {pages} {855} (\bibinfo {year} {2015})}\BibitemShut {NoStop}%
\bibitem [{\citenamefont {Georges}\ \emph {et~al.}(1996)\citenamefont
  {Georges}, \citenamefont {Kotliar}, \citenamefont {Krauth},\ and\
  \citenamefont {Rozenberg}}]{RevModPhys.68.13}%
  \BibitemOpen
  \bibfield  {author} {\bibinfo {author} {\bibfnamefont {A.}~\bibnamefont
  {Georges}}, \bibinfo {author} {\bibfnamefont {G.}~\bibnamefont {Kotliar}},
  \bibinfo {author} {\bibfnamefont {W.}~\bibnamefont {Krauth}}, \ and\ \bibinfo
  {author} {\bibfnamefont {M.~J.}\ \bibnamefont {Rozenberg}},\ }\href {\doibase
  10.1103/RevModPhys.68.13} {\bibfield  {journal} {\bibinfo  {journal} {Rev.
  Mod. Phys.}\ }\textbf {\bibinfo {volume} {68}},\ \bibinfo {pages} {13}
  (\bibinfo {year} {1996})}\BibitemShut {NoStop}%
\bibitem [{\citenamefont {Abanov}\ and\ \citenamefont
  {Chubukov}(2000)}]{PhysRevLett.84.5608}%
  \BibitemOpen
  \bibfield  {author} {\bibinfo {author} {\bibfnamefont {A.}~\bibnamefont
  {Abanov}}\ and\ \bibinfo {author} {\bibfnamefont {A.~V.}\ \bibnamefont
  {Chubukov}},\ }\href {\doibase 10.1103/PhysRevLett.84.5608} {\bibfield
  {journal} {\bibinfo  {journal} {Phys. Rev. Lett.}\ }\textbf {\bibinfo
  {volume} {84}},\ \bibinfo {pages} {5608} (\bibinfo {year}
  {2000})}\BibitemShut {NoStop}%
\bibitem [{\citenamefont {Abanov}\ \emph {et~al.}(2003)\citenamefont {Abanov},
  \citenamefont {Chubukov},\ and\ \citenamefont
  {Schmalian}}]{spin_fermion_abanov}%
  \BibitemOpen
  \bibfield  {author} {\bibinfo {author} {\bibfnamefont {A.}~\bibnamefont
  {Abanov}}, \bibinfo {author} {\bibfnamefont {A.~V.}\ \bibnamefont
  {Chubukov}}, \ and\ \bibinfo {author} {\bibfnamefont {J.}~\bibnamefont
  {Schmalian}},\ }\href {\doibase 10.1080/0001873021000057123} {\bibfield
  {journal} {\bibinfo  {journal} {Advances in Physics}\ }\textbf {\bibinfo
  {volume} {52}},\ \bibinfo {pages} {119} (\bibinfo {year} {2003})}\BibitemShut
  {NoStop}%
\bibitem [{\citenamefont {Werner}\ \emph {et~al.}(2006)\citenamefont {Werner},
  \citenamefont {Comanac}, \citenamefont {de' Medici}, \citenamefont {Troyer},\
  and\ \citenamefont {Millis}}]{PhysRevLett.97.076405}%
  \BibitemOpen
  \bibfield  {author} {\bibinfo {author} {\bibfnamefont {P.}~\bibnamefont
  {Werner}}, \bibinfo {author} {\bibfnamefont {A.}~\bibnamefont {Comanac}},
  \bibinfo {author} {\bibfnamefont {L.}~\bibnamefont {de' Medici}}, \bibinfo
  {author} {\bibfnamefont {M.}~\bibnamefont {Troyer}}, \ and\ \bibinfo {author}
  {\bibfnamefont {A.~J.}\ \bibnamefont {Millis}},\ }\href {\doibase
  10.1103/PhysRevLett.97.076405} {\bibfield  {journal} {\bibinfo  {journal}
  {Phys. Rev. Lett.}\ }\textbf {\bibinfo {volume} {97}},\ \bibinfo {pages}
  {076405} (\bibinfo {year} {2006})}\BibitemShut {NoStop}%
\bibitem [{\citenamefont {Haule}(2007)}]{PhysRevB.75.155113}%
  \BibitemOpen
  \bibfield  {author} {\bibinfo {author} {\bibfnamefont {K.}~\bibnamefont
  {Haule}},\ }\href {\doibase 10.1103/PhysRevB.75.155113} {\bibfield  {journal}
  {\bibinfo  {journal} {Phys. Rev. B}\ }\textbf {\bibinfo {volume} {75}},\
  \bibinfo {pages} {155113} (\bibinfo {year} {2007})}\BibitemShut {NoStop}%
\bibitem [{\citenamefont {Park}\ \emph {et~al.}(2008)\citenamefont {Park},
  \citenamefont {Haule},\ and\ \citenamefont
  {Kotliar}}]{PhysRevLett.101.186403}%
  \BibitemOpen
  \bibfield  {author} {\bibinfo {author} {\bibfnamefont {H.}~\bibnamefont
  {Park}}, \bibinfo {author} {\bibfnamefont {K.}~\bibnamefont {Haule}}, \ and\
  \bibinfo {author} {\bibfnamefont {G.}~\bibnamefont {Kotliar}},\ }\href
  {\doibase 10.1103/PhysRevLett.101.186403} {\bibfield  {journal} {\bibinfo
  {journal} {Phys. Rev. Lett.}\ }\textbf {\bibinfo {volume} {101}},\ \bibinfo
  {pages} {186403} (\bibinfo {year} {2008})}\BibitemShut {NoStop}%
\bibitem [{\citenamefont {Toschi}\ \emph {et~al.}(2007)\citenamefont {Toschi},
  \citenamefont {Katanin},\ and\ \citenamefont {Held}}]{PhysRevB.75.045118}%
  \BibitemOpen
  \bibfield  {author} {\bibinfo {author} {\bibfnamefont {A.}~\bibnamefont
  {Toschi}}, \bibinfo {author} {\bibfnamefont {A.~A.}\ \bibnamefont {Katanin}},
  \ and\ \bibinfo {author} {\bibfnamefont {K.}~\bibnamefont {Held}},\ }\href
  {\doibase 10.1103/PhysRevB.75.045118} {\bibfield  {journal} {\bibinfo
  {journal} {Phys. Rev. B}\ }\textbf {\bibinfo {volume} {75}},\ \bibinfo
  {pages} {045118} (\bibinfo {year} {2007})}\BibitemShut {NoStop}%
\bibitem [{\citenamefont {Rubtsov}\ \emph {et~al.}(2008)\citenamefont
  {Rubtsov}, \citenamefont {Katsnelson},\ and\ \citenamefont
  {Lichtenstein}}]{PhysRevB.77.033101}%
  \BibitemOpen
  \bibfield  {author} {\bibinfo {author} {\bibfnamefont {A.~N.}\ \bibnamefont
  {Rubtsov}}, \bibinfo {author} {\bibfnamefont {M.~I.}\ \bibnamefont
  {Katsnelson}}, \ and\ \bibinfo {author} {\bibfnamefont {A.~I.}\ \bibnamefont
  {Lichtenstein}},\ }\href {\doibase 10.1103/PhysRevB.77.033101} {\bibfield
  {journal} {\bibinfo  {journal} {Phys. Rev. B}\ }\textbf {\bibinfo {volume}
  {77}},\ \bibinfo {pages} {033101} (\bibinfo {year} {2008})}\BibitemShut
  {NoStop}%
\bibitem [{\citenamefont {Xu}\ \emph {et~al.}(2013)\citenamefont {Xu},
  \citenamefont {Haule},\ and\ \citenamefont
  {Kotliar}}]{PhysRevLett.111.036401}%
  \BibitemOpen
  \bibfield  {author} {\bibinfo {author} {\bibfnamefont {W.}~\bibnamefont
  {Xu}}, \bibinfo {author} {\bibfnamefont {K.}~\bibnamefont {Haule}}, \ and\
  \bibinfo {author} {\bibfnamefont {G.}~\bibnamefont {Kotliar}},\ }\href
  {\doibase 10.1103/PhysRevLett.111.036401} {\bibfield  {journal} {\bibinfo
  {journal} {Phys. Rev. Lett.}\ }\textbf {\bibinfo {volume} {111}},\ \bibinfo
  {pages} {036401} (\bibinfo {year} {2013})}\BibitemShut {NoStop}%
\bibitem [{\citenamefont {Deng}\ \emph {et~al.}(2013)\citenamefont {Deng},
  \citenamefont {Mravlje}, \citenamefont {\ifmmode~\check{Z}\else
  \v{Z}\fi{}itko}, \citenamefont {Ferrero}, \citenamefont {Kotliar},\ and\
  \citenamefont {Georges}}]{PhysRevLett.110.086401}%
  \BibitemOpen
  \bibfield  {author} {\bibinfo {author} {\bibfnamefont {X.}~\bibnamefont
  {Deng}}, \bibinfo {author} {\bibfnamefont {J.}~\bibnamefont {Mravlje}},
  \bibinfo {author} {\bibfnamefont {R.}~\bibnamefont {\ifmmode~\check{Z}\else
  \v{Z}\fi{}itko}}, \bibinfo {author} {\bibfnamefont {M.}~\bibnamefont
  {Ferrero}}, \bibinfo {author} {\bibfnamefont {G.}~\bibnamefont {Kotliar}}, \
  and\ \bibinfo {author} {\bibfnamefont {A.}~\bibnamefont {Georges}},\ }\href
  {\doibase 10.1103/PhysRevLett.110.086401} {\bibfield  {journal} {\bibinfo
  {journal} {Phys. Rev. Lett.}\ }\textbf {\bibinfo {volume} {110}},\ \bibinfo
  {pages} {086401} (\bibinfo {year} {2013})}\BibitemShut {NoStop}%
\bibitem [{\citenamefont {Deng}\ \emph {et~al.}(2014)\citenamefont {Deng},
  \citenamefont {Sternbach}, \citenamefont {Haule}, \citenamefont {Basov},\
  and\ \citenamefont {Kotliar}}]{PhysRevLett.113.246404}%
  \BibitemOpen
  \bibfield  {author} {\bibinfo {author} {\bibfnamefont {X.}~\bibnamefont
  {Deng}}, \bibinfo {author} {\bibfnamefont {A.}~\bibnamefont {Sternbach}},
  \bibinfo {author} {\bibfnamefont {K.}~\bibnamefont {Haule}}, \bibinfo
  {author} {\bibfnamefont {D.~N.}\ \bibnamefont {Basov}}, \ and\ \bibinfo
  {author} {\bibfnamefont {G.}~\bibnamefont {Kotliar}},\ }\href {\doibase
  10.1103/PhysRevLett.113.246404} {\bibfield  {journal} {\bibinfo  {journal}
  {Phys. Rev. Lett.}\ }\textbf {\bibinfo {volume} {113}},\ \bibinfo {pages}
  {246404} (\bibinfo {year} {2014})}\BibitemShut {NoStop}%
\bibitem [{\citenamefont {Mai}\ \emph {et~al.}(2016)\citenamefont {Mai},
  \citenamefont {Krishna-murthy},\ and\ \citenamefont {Shastry}}]{Mai2016}%
  \BibitemOpen
  \bibfield  {author} {\bibinfo {author} {\bibfnamefont {P.}~\bibnamefont
  {Mai}}, \bibinfo {author} {\bibfnamefont {H.}~\bibnamefont {Krishna-murthy}},
  \ and\ \bibinfo {author} {\bibfnamefont {B.~S.}\ \bibnamefont {Shastry}},\
  }\href {\doibase http://dx.doi.org/10.1016/j.aop.2016.03.011} {\bibfield
  {journal} {\bibinfo  {journal} {Annals of Physics}\ ,\ } (\bibinfo {year}
  {2016})},\ \bibinfo {note} {in press}\BibitemShut {NoStop}%
\bibitem [{\citenamefont {Graf}\ \emph {et~al.}(2007)\citenamefont {Graf},
  \citenamefont {Gweon}, \citenamefont {McElroy}, \citenamefont {Zhou},
  \citenamefont {Jozwiak}, \citenamefont {Rotenberg}, \citenamefont {Bill},
  \citenamefont {Sasagawa}, \citenamefont {Eisaki}, \citenamefont {Uchida},
  \citenamefont {Takagi}, \citenamefont {Lee},\ and\ \citenamefont
  {Lanzara}}]{PhysRevLett.98.067004}%
  \BibitemOpen
  \bibfield  {author} {\bibinfo {author} {\bibfnamefont {J.}~\bibnamefont
  {Graf}}, \bibinfo {author} {\bibfnamefont {G.-H.}\ \bibnamefont {Gweon}},
  \bibinfo {author} {\bibfnamefont {K.}~\bibnamefont {McElroy}}, \bibinfo
  {author} {\bibfnamefont {S.~Y.}\ \bibnamefont {Zhou}}, \bibinfo {author}
  {\bibfnamefont {C.}~\bibnamefont {Jozwiak}}, \bibinfo {author} {\bibfnamefont
  {E.}~\bibnamefont {Rotenberg}}, \bibinfo {author} {\bibfnamefont
  {A.}~\bibnamefont {Bill}}, \bibinfo {author} {\bibfnamefont {T.}~\bibnamefont
  {Sasagawa}}, \bibinfo {author} {\bibfnamefont {H.}~\bibnamefont {Eisaki}},
  \bibinfo {author} {\bibfnamefont {S.}~\bibnamefont {Uchida}}, \bibinfo
  {author} {\bibfnamefont {H.}~\bibnamefont {Takagi}}, \bibinfo {author}
  {\bibfnamefont {D.-H.}\ \bibnamefont {Lee}}, \ and\ \bibinfo {author}
  {\bibfnamefont {A.}~\bibnamefont {Lanzara}},\ }\href {\doibase
  10.1103/PhysRevLett.98.067004} {\bibfield  {journal} {\bibinfo  {journal}
  {Phys. Rev. Lett.}\ }\textbf {\bibinfo {volume} {98}},\ \bibinfo {pages}
  {067004} (\bibinfo {year} {2007})}\BibitemShut {NoStop}%
\bibitem [{\citenamefont {Inosov}\ \emph {et~al.}(2007)\citenamefont {Inosov},
  \citenamefont {Fink}, \citenamefont {Kordyuk}, \citenamefont {Borisenko},
  \citenamefont {Zabolotnyy}, \citenamefont {Schuster}, \citenamefont
  {Knupfer}, \citenamefont {B\"uchner}, \citenamefont {Follath}, \citenamefont
  {D\"urr}, \citenamefont {Eberhardt}, \citenamefont {Hinkov}, \citenamefont
  {Keimer},\ and\ \citenamefont {Berger}}]{PhysRevLett.99.237002}%
  \BibitemOpen
  \bibfield  {author} {\bibinfo {author} {\bibfnamefont {D.~S.}\ \bibnamefont
  {Inosov}}, \bibinfo {author} {\bibfnamefont {J.}~\bibnamefont {Fink}},
  \bibinfo {author} {\bibfnamefont {A.~A.}\ \bibnamefont {Kordyuk}}, \bibinfo
  {author} {\bibfnamefont {S.~V.}\ \bibnamefont {Borisenko}}, \bibinfo {author}
  {\bibfnamefont {V.~B.}\ \bibnamefont {Zabolotnyy}}, \bibinfo {author}
  {\bibfnamefont {R.}~\bibnamefont {Schuster}}, \bibinfo {author}
  {\bibfnamefont {M.}~\bibnamefont {Knupfer}}, \bibinfo {author} {\bibfnamefont
  {B.}~\bibnamefont {B\"uchner}}, \bibinfo {author} {\bibfnamefont
  {R.}~\bibnamefont {Follath}}, \bibinfo {author} {\bibfnamefont {H.~A.}\
  \bibnamefont {D\"urr}}, \bibinfo {author} {\bibfnamefont {W.}~\bibnamefont
  {Eberhardt}}, \bibinfo {author} {\bibfnamefont {V.}~\bibnamefont {Hinkov}},
  \bibinfo {author} {\bibfnamefont {B.}~\bibnamefont {Keimer}}, \ and\ \bibinfo
  {author} {\bibfnamefont {H.}~\bibnamefont {Berger}},\ }\href {\doibase
  10.1103/PhysRevLett.99.237002} {\bibfield  {journal} {\bibinfo  {journal}
  {Phys. Rev. Lett.}\ }\textbf {\bibinfo {volume} {99}},\ \bibinfo {pages}
  {237002} (\bibinfo {year} {2007})}\BibitemShut {NoStop}%
\bibitem [{\citenamefont {Iwasawa}\ \emph {et~al.}(2012)\citenamefont
  {Iwasawa}, \citenamefont {Yoshida}, \citenamefont {Hase}, \citenamefont
  {Shimada}, \citenamefont {Namatame}, \citenamefont {Taniguchi},\ and\
  \citenamefont {Aiura}}]{PhysRevLett.109.066404}%
  \BibitemOpen
  \bibfield  {author} {\bibinfo {author} {\bibfnamefont {H.}~\bibnamefont
  {Iwasawa}}, \bibinfo {author} {\bibfnamefont {Y.}~\bibnamefont {Yoshida}},
  \bibinfo {author} {\bibfnamefont {I.}~\bibnamefont {Hase}}, \bibinfo {author}
  {\bibfnamefont {K.}~\bibnamefont {Shimada}}, \bibinfo {author} {\bibfnamefont
  {H.}~\bibnamefont {Namatame}}, \bibinfo {author} {\bibfnamefont
  {M.}~\bibnamefont {Taniguchi}}, \ and\ \bibinfo {author} {\bibfnamefont
  {Y.}~\bibnamefont {Aiura}},\ }\href {\doibase 10.1103/PhysRevLett.109.066404}
  {\bibfield  {journal} {\bibinfo  {journal} {Phys. Rev. Lett.}\ }\textbf
  {\bibinfo {volume} {109}},\ \bibinfo {pages} {066404} (\bibinfo {year}
  {2012})}\BibitemShut {NoStop}%
\bibitem [{\citenamefont {Das}\ \emph {et~al.}(2012)\citenamefont {Das},
  \citenamefont {Durakiewicz}, \citenamefont {Zhu}, \citenamefont {Joyce},
  \citenamefont {Sarrao},\ and\ \citenamefont {Graf}}]{PhysRevX.2.041012}%
  \BibitemOpen
  \bibfield  {author} {\bibinfo {author} {\bibfnamefont {T.}~\bibnamefont
  {Das}}, \bibinfo {author} {\bibfnamefont {T.}~\bibnamefont {Durakiewicz}},
  \bibinfo {author} {\bibfnamefont {J.-X.}\ \bibnamefont {Zhu}}, \bibinfo
  {author} {\bibfnamefont {J.~J.}\ \bibnamefont {Joyce}}, \bibinfo {author}
  {\bibfnamefont {J.~L.}\ \bibnamefont {Sarrao}}, \ and\ \bibinfo {author}
  {\bibfnamefont {M.~J.}\ \bibnamefont {Graf}},\ }\href {\doibase
  10.1103/PhysRevX.2.041012} {\bibfield  {journal} {\bibinfo  {journal} {Phys.
  Rev. X}\ }\textbf {\bibinfo {volume} {2}},\ \bibinfo {pages} {041012}
  (\bibinfo {year} {2012})}\BibitemShut {NoStop}%
\bibitem [{\citenamefont {Hubbard}(1963)}]{Hubbard238}%
  \BibitemOpen
  \bibfield  {author} {\bibinfo {author} {\bibfnamefont {J.}~\bibnamefont
  {Hubbard}},\ }\href {\doibase 10.1098/rspa.1963.0204} {\bibfield  {journal}
  {\bibinfo  {journal} {Proceedings of the Royal Society of London A:
  Mathematical, Physical and Engineering Sciences}\ }\textbf {\bibinfo {volume}
  {276}},\ \bibinfo {pages} {238} (\bibinfo {year} {1963})}\BibitemShut
  {NoStop}%
\bibitem [{\citenamefont {Macridin}\ \emph {et~al.}(2007)\citenamefont
  {Macridin}, \citenamefont {Jarrell}, \citenamefont {Maier},\ and\
  \citenamefont {Scalapino}}]{PhysRevLett.99.237001}%
  \BibitemOpen
  \bibfield  {author} {\bibinfo {author} {\bibfnamefont {A.}~\bibnamefont
  {Macridin}}, \bibinfo {author} {\bibfnamefont {M.}~\bibnamefont {Jarrell}},
  \bibinfo {author} {\bibfnamefont {T.}~\bibnamefont {Maier}}, \ and\ \bibinfo
  {author} {\bibfnamefont {D.~J.}\ \bibnamefont {Scalapino}},\ }\href {\doibase
  10.1103/PhysRevLett.99.237001} {\bibfield  {journal} {\bibinfo  {journal}
  {Phys. Rev. Lett.}\ }\textbf {\bibinfo {volume} {99}},\ \bibinfo {pages}
  {237001} (\bibinfo {year} {2007})}\BibitemShut {NoStop}%
\bibitem [{\citenamefont {Weber}\ \emph {et~al.}(2008)\citenamefont {Weber},
  \citenamefont {Haule},\ and\ \citenamefont {Kotliar}}]{PhysRevB.78.134519}%
  \BibitemOpen
  \bibfield  {author} {\bibinfo {author} {\bibfnamefont {C.}~\bibnamefont
  {Weber}}, \bibinfo {author} {\bibfnamefont {K.}~\bibnamefont {Haule}}, \ and\
  \bibinfo {author} {\bibfnamefont {G.}~\bibnamefont {Kotliar}},\ }\href
  {\doibase 10.1103/PhysRevB.78.134519} {\bibfield  {journal} {\bibinfo
  {journal} {Phys. Rev. B}\ }\textbf {\bibinfo {volume} {78}},\ \bibinfo
  {pages} {134519} (\bibinfo {year} {2008})}\BibitemShut {NoStop}%
\bibitem [{\citenamefont {Yin}\ \emph {et~al.}(2008)\citenamefont {Yin},
  \citenamefont {Gordienko}, \citenamefont {Wan},\ and\ \citenamefont
  {Savrasov}}]{PhysRevLett.100.066406}%
  \BibitemOpen
  \bibfield  {author} {\bibinfo {author} {\bibfnamefont {Q.}~\bibnamefont
  {Yin}}, \bibinfo {author} {\bibfnamefont {A.}~\bibnamefont {Gordienko}},
  \bibinfo {author} {\bibfnamefont {X.}~\bibnamefont {Wan}}, \ and\ \bibinfo
  {author} {\bibfnamefont {S.~Y.}\ \bibnamefont {Savrasov}},\ }\href {\doibase
  10.1103/PhysRevLett.100.066406} {\bibfield  {journal} {\bibinfo  {journal}
  {Phys. Rev. Lett.}\ }\textbf {\bibinfo {volume} {100}},\ \bibinfo {pages}
  {066406} (\bibinfo {year} {2008})}\BibitemShut {NoStop}%
\bibitem [{\citenamefont {Chakraborty}\ \emph {et~al.}(2008)\citenamefont
  {Chakraborty}, \citenamefont {Galanakis},\ and\ \citenamefont
  {Phillips}}]{PhysRevB.78.212504}%
  \BibitemOpen
  \bibfield  {author} {\bibinfo {author} {\bibfnamefont {S.}~\bibnamefont
  {Chakraborty}}, \bibinfo {author} {\bibfnamefont {D.}~\bibnamefont
  {Galanakis}}, \ and\ \bibinfo {author} {\bibfnamefont {P.}~\bibnamefont
  {Phillips}},\ }\href {\doibase 10.1103/PhysRevB.78.212504} {\bibfield
  {journal} {\bibinfo  {journal} {Phys. Rev. B}\ }\textbf {\bibinfo {volume}
  {78}},\ \bibinfo {pages} {212504} (\bibinfo {year} {2008})}\BibitemShut
  {NoStop}%
\bibitem [{\citenamefont {Sakai}\ \emph {et~al.}(2010)\citenamefont {Sakai},
  \citenamefont {Motome},\ and\ \citenamefont {Imada}}]{PhysRevB.82.134505}%
  \BibitemOpen
  \bibfield  {author} {\bibinfo {author} {\bibfnamefont {S.}~\bibnamefont
  {Sakai}}, \bibinfo {author} {\bibfnamefont {Y.}~\bibnamefont {Motome}}, \
  and\ \bibinfo {author} {\bibfnamefont {M.}~\bibnamefont {Imada}},\ }\href
  {\doibase 10.1103/PhysRevB.82.134505} {\bibfield  {journal} {\bibinfo
  {journal} {Phys. Rev. B}\ }\textbf {\bibinfo {volume} {82}},\ \bibinfo
  {pages} {134505} (\bibinfo {year} {2010})}\BibitemShut {NoStop}%
\bibitem [{\citenamefont {Fuchs}\ \emph {et~al.}(2011)\citenamefont {Fuchs},
  \citenamefont {Gull}, \citenamefont {Troyer}, \citenamefont {Jarrell},\ and\
  \citenamefont {Pruschke}}]{PhysRevB.83.235113}%
  \BibitemOpen
  \bibfield  {author} {\bibinfo {author} {\bibfnamefont {S.}~\bibnamefont
  {Fuchs}}, \bibinfo {author} {\bibfnamefont {E.}~\bibnamefont {Gull}},
  \bibinfo {author} {\bibfnamefont {M.}~\bibnamefont {Troyer}}, \bibinfo
  {author} {\bibfnamefont {M.}~\bibnamefont {Jarrell}}, \ and\ \bibinfo
  {author} {\bibfnamefont {T.}~\bibnamefont {Pruschke}},\ }\href {\doibase
  10.1103/PhysRevB.83.235113} {\bibfield  {journal} {\bibinfo  {journal} {Phys.
  Rev. B}\ }\textbf {\bibinfo {volume} {83}},\ \bibinfo {pages} {235113}
  (\bibinfo {year} {2011})}\BibitemShut {NoStop}%
\bibitem [{\citenamefont {Hinkov}\ \emph {et~al.}(2007)\citenamefont {Hinkov},
  \citenamefont {Bourges}, \citenamefont {Pailhes}, \citenamefont {Sidis},
  \citenamefont {Ivanov}, \citenamefont {Frost}, \citenamefont {Perring},
  \citenamefont {Lin}, \citenamefont {Chen},\ and\ \citenamefont
  {Keimer}}]{hinkov2007spin}%
  \BibitemOpen
  \bibfield  {author} {\bibinfo {author} {\bibfnamefont {V.}~\bibnamefont
  {Hinkov}}, \bibinfo {author} {\bibfnamefont {P.}~\bibnamefont {Bourges}},
  \bibinfo {author} {\bibfnamefont {S.}~\bibnamefont {Pailhes}}, \bibinfo
  {author} {\bibfnamefont {Y.}~\bibnamefont {Sidis}}, \bibinfo {author}
  {\bibfnamefont {A.}~\bibnamefont {Ivanov}}, \bibinfo {author} {\bibfnamefont
  {C.}~\bibnamefont {Frost}}, \bibinfo {author} {\bibfnamefont
  {T.}~\bibnamefont {Perring}}, \bibinfo {author} {\bibfnamefont
  {C.}~\bibnamefont {Lin}}, \bibinfo {author} {\bibfnamefont {D.}~\bibnamefont
  {Chen}}, \ and\ \bibinfo {author} {\bibfnamefont {B.}~\bibnamefont
  {Keimer}},\ }\href@noop {} {\bibfield  {journal} {\bibinfo  {journal} {Nature
  Physics}\ }\textbf {\bibinfo {volume} {3}},\ \bibinfo {pages} {780} (\bibinfo
  {year} {2007})}\BibitemShut {NoStop}%
\bibitem [{\citenamefont {Vignolle}\ \emph {et~al.}(2007)\citenamefont
  {Vignolle}, \citenamefont {Hayden}, \citenamefont {McMorrow}, \citenamefont
  {R{\o}nnow}, \citenamefont {Lake}, \citenamefont {Frost},\ and\ \citenamefont
  {Perring}}]{vignolle2007two}%
  \BibitemOpen
  \bibfield  {author} {\bibinfo {author} {\bibfnamefont {B.}~\bibnamefont
  {Vignolle}}, \bibinfo {author} {\bibfnamefont {S.}~\bibnamefont {Hayden}},
  \bibinfo {author} {\bibfnamefont {D.}~\bibnamefont {McMorrow}}, \bibinfo
  {author} {\bibfnamefont {H.}~\bibnamefont {R{\o}nnow}}, \bibinfo {author}
  {\bibfnamefont {B.}~\bibnamefont {Lake}}, \bibinfo {author} {\bibfnamefont
  {C.}~\bibnamefont {Frost}}, \ and\ \bibinfo {author} {\bibfnamefont
  {T.}~\bibnamefont {Perring}},\ }\href@noop {} {\bibfield  {journal} {\bibinfo
   {journal} {Nature Physics}\ }\textbf {\bibinfo {volume} {3}},\ \bibinfo
  {pages} {163} (\bibinfo {year} {2007})}\BibitemShut {NoStop}%
\bibitem [{\citenamefont {Reznik}\ \emph {et~al.}(2008)\citenamefont {Reznik},
  \citenamefont {Ismer}, \citenamefont {Eremin}, \citenamefont {Pintschovius},
  \citenamefont {Wolf}, \citenamefont {Arai}, \citenamefont {Endoh},
  \citenamefont {Masui},\ and\ \citenamefont {Tajima}}]{PhysRevB.78.132503}%
  \BibitemOpen
  \bibfield  {author} {\bibinfo {author} {\bibfnamefont {D.}~\bibnamefont
  {Reznik}}, \bibinfo {author} {\bibfnamefont {J.-P.}\ \bibnamefont {Ismer}},
  \bibinfo {author} {\bibfnamefont {I.}~\bibnamefont {Eremin}}, \bibinfo
  {author} {\bibfnamefont {L.}~\bibnamefont {Pintschovius}}, \bibinfo {author}
  {\bibfnamefont {T.}~\bibnamefont {Wolf}}, \bibinfo {author} {\bibfnamefont
  {M.}~\bibnamefont {Arai}}, \bibinfo {author} {\bibfnamefont {Y.}~\bibnamefont
  {Endoh}}, \bibinfo {author} {\bibfnamefont {T.}~\bibnamefont {Masui}}, \ and\
  \bibinfo {author} {\bibfnamefont {S.}~\bibnamefont {Tajima}},\ }\href
  {\doibase 10.1103/PhysRevB.78.132503} {\bibfield  {journal} {\bibinfo
  {journal} {Phys. Rev. B}\ }\textbf {\bibinfo {volume} {78}},\ \bibinfo
  {pages} {132503} (\bibinfo {year} {2008})}\BibitemShut {NoStop}%
\bibitem [{\citenamefont {Xu}\ \emph {et~al.}(2009)\citenamefont {Xu},
  \citenamefont {Gu}, \citenamefont {H{\"u}cker}, \citenamefont {Fauqu{\'e}},
  \citenamefont {Perring}, \citenamefont {Regnault},\ and\ \citenamefont
  {Tranquada}}]{xu2009testing}%
  \BibitemOpen
  \bibfield  {author} {\bibinfo {author} {\bibfnamefont {G.}~\bibnamefont
  {Xu}}, \bibinfo {author} {\bibfnamefont {G.}~\bibnamefont {Gu}}, \bibinfo
  {author} {\bibfnamefont {M.}~\bibnamefont {H{\"u}cker}}, \bibinfo {author}
  {\bibfnamefont {B.}~\bibnamefont {Fauqu{\'e}}}, \bibinfo {author}
  {\bibfnamefont {T.}~\bibnamefont {Perring}}, \bibinfo {author} {\bibfnamefont
  {L.}~\bibnamefont {Regnault}}, \ and\ \bibinfo {author} {\bibfnamefont
  {J.}~\bibnamefont {Tranquada}},\ }\href@noop {} {\bibfield  {journal}
  {\bibinfo  {journal} {Nature Physics}\ }\textbf {\bibinfo {volume} {5}},\
  \bibinfo {pages} {642} (\bibinfo {year} {2009})}\BibitemShut {NoStop}%
\bibitem [{\citenamefont {Le~Tacon}\ \emph {et~al.}(2011)\citenamefont
  {Le~Tacon}, \citenamefont {Ghiringhelli}, \citenamefont {Chaloupka},
  \citenamefont {Sala}, \citenamefont {Hinkov}, \citenamefont {Haverkort},
  \citenamefont {Minola}, \citenamefont {Bakr}, \citenamefont {Zhou},
  \citenamefont {Blanco-Canosa} \emph {et~al.}}]{le2011intense}%
  \BibitemOpen
  \bibfield  {author} {\bibinfo {author} {\bibfnamefont {M.}~\bibnamefont
  {Le~Tacon}}, \bibinfo {author} {\bibfnamefont {G.}~\bibnamefont
  {Ghiringhelli}}, \bibinfo {author} {\bibfnamefont {J.}~\bibnamefont
  {Chaloupka}}, \bibinfo {author} {\bibfnamefont {M.~M.}\ \bibnamefont {Sala}},
  \bibinfo {author} {\bibfnamefont {V.}~\bibnamefont {Hinkov}}, \bibinfo
  {author} {\bibfnamefont {M.}~\bibnamefont {Haverkort}}, \bibinfo {author}
  {\bibfnamefont {M.}~\bibnamefont {Minola}}, \bibinfo {author} {\bibfnamefont
  {M.}~\bibnamefont {Bakr}}, \bibinfo {author} {\bibfnamefont {K.}~\bibnamefont
  {Zhou}}, \bibinfo {author} {\bibfnamefont {S.}~\bibnamefont {Blanco-Canosa}},
   \emph {et~al.},\ }\href@noop {} {\bibfield  {journal} {\bibinfo  {journal}
  {Nature Physics}\ }\textbf {\bibinfo {volume} {7}},\ \bibinfo {pages} {725}
  (\bibinfo {year} {2011})}\BibitemShut {NoStop}%
\bibitem [{\citenamefont {Jia}\ \emph {et~al.}(2014)\citenamefont {Jia},
  \citenamefont {Nowadnick}, \citenamefont {Wohlfeld}, \citenamefont {Kung},
  \citenamefont {Chen}, \citenamefont {Johnston}, \citenamefont {Tohyama},
  \citenamefont {Moritz},\ and\ \citenamefont {Devereaux}}]{jia2014persistent}%
  \BibitemOpen
  \bibfield  {author} {\bibinfo {author} {\bibfnamefont {C.}~\bibnamefont
  {Jia}}, \bibinfo {author} {\bibfnamefont {E.}~\bibnamefont {Nowadnick}},
  \bibinfo {author} {\bibfnamefont {K.}~\bibnamefont {Wohlfeld}}, \bibinfo
  {author} {\bibfnamefont {Y.}~\bibnamefont {Kung}}, \bibinfo {author}
  {\bibfnamefont {C.-C.}\ \bibnamefont {Chen}}, \bibinfo {author}
  {\bibfnamefont {S.}~\bibnamefont {Johnston}}, \bibinfo {author}
  {\bibfnamefont {T.}~\bibnamefont {Tohyama}}, \bibinfo {author} {\bibfnamefont
  {B.}~\bibnamefont {Moritz}}, \ and\ \bibinfo {author} {\bibfnamefont
  {T.}~\bibnamefont {Devereaux}},\ }\href@noop {} {\bibfield  {journal}
  {\bibinfo  {journal} {Nature communications}\ }\textbf {\bibinfo {volume}
  {5}} (\bibinfo {year} {2014})}\BibitemShut {NoStop}%
\bibitem [{\citenamefont {Rubtsov}\ \emph {et~al.}(2009)\citenamefont
  {Rubtsov}, \citenamefont {Katsnelson}, \citenamefont {Lichtenstein},\ and\
  \citenamefont {Georges}}]{PhysRevB.79.045133}%
  \BibitemOpen
  \bibfield  {author} {\bibinfo {author} {\bibfnamefont {A.~N.}\ \bibnamefont
  {Rubtsov}}, \bibinfo {author} {\bibfnamefont {M.~I.}\ \bibnamefont
  {Katsnelson}}, \bibinfo {author} {\bibfnamefont {A.~I.}\ \bibnamefont
  {Lichtenstein}}, \ and\ \bibinfo {author} {\bibfnamefont {A.}~\bibnamefont
  {Georges}},\ }\href {\doibase 10.1103/PhysRevB.79.045133} {\bibfield
  {journal} {\bibinfo  {journal} {Phys. Rev. B}\ }\textbf {\bibinfo {volume}
  {79}},\ \bibinfo {pages} {045133} (\bibinfo {year} {2009})}\BibitemShut
  {NoStop}%
\end{thebibliography}%

\end{document}